# Innovative activities of Activision Blizzard: A patent network analysis

**Artur F. Tomeczek**

*SGH Warsaw School of Economics*

*artur.tomeczek@sgh.waw.pl*

*https://orcid.org/0000-0003-4888-2535*

**Abstract:** Microsoft's acquisition of Activision Blizzard valued at $68.7 billion ($95 per share) has drastically altered the landscape of the video game industry. At the time of the takeover, the intellectual properties of Activision Blizzard included World of Warcraft, Diablo, Hearthstone, StarCraft, Overwatch, Battle.net, Candy Crush Saga, and Call of Duty. This article aims to explore the patenting activity of Activision Blizzard between 2008 (the original merger) and 2023 (the Microsoft acquisition). Four IPC code co-occurrence networks (co-classification maps) are constructed and analyzed based on the patent data downloaded from the WIPO Patentscope database. International Patent Classification (IPC) codes are a language agnostic system for the classification of patents. When multiple IPC codes co-occur in a patent, it shows that the technologies are connected. These relationships can be used for patent mapping. The analysis identifies the prolific and bridging technologies of Activision Blizzard and explores its synergistic role as a subsidiary of Microsoft Corporation.

## 1. Introduction

Microsoft's acquisition of Activision Blizzard has drastically altered the landscape of the video game industry. The all-cash transaction valued at $68.7 billion ($95 per share) was initially announced on January 18, 2022 [1]. The record-breaking move generated both major excitement and significant controversy, including anti-consumerist issues raised by competitors, primarily Sony. Due to potential monopolistic concerns, a series of regulatory investigations were launched by the European Commission (European Union), the Federal Trade Commission (United States), as well as the Competition and Markets Authority (United Kingdom) [2], [3], [4], [5], [6]. The long and arduous acquisition process was completed on October 13, 2023 [7], [8]. The total purchase price was $75.4 billion, and the valuation of Activision Blizzard's assets included cash and cash equivalents of $13 billion, intangible assets of $22 billion, and goodwill of $51 billion [8]. Several studies have explored various business and legal aspects of the initial phase of the acquisition [9], [10], [11], [12].

Patents are used by firms to protect their intellectual property. International Patent Classification (IPC) codes are a language agnostic system for the classification of patents [13], [14]. When multiple IPC codes co-occur (are co-classified) in a patent, it shows that the technologies are connected. The complex relationships between the codes can be modeled using network methodology. Such networks can present information contained in a corpus of publications in the form of a knowledge map. Popular in bibliometric analysis of scientific articles, keyword co-occurrence networks are an established method of knowledge mapping [15]. When applied to patent publications, the construction of IPC code co-occurrence networks is similar to their keyword-based counterparts, and they both can be analyzed using standard network centrality measures.

This article aims to explore the patenting activity of Activision Blizzard between 2008 (the original merger) and 2023 (the Microsoft acquisition). Four IPC code co-occurrence networks (co-classification maps) are constructed and analyzed based on Activision Blizzard patent data downloaded from the WIPO Patentscope database. Nodes in networks represent IPC codes (technologies), and edges that connect them represent the co-occurrence of IPC codes in a patent. The analysis identifies the prolific and bridging technologies of Activision Blizzard and explores its synergistic role as a subsidiary of Microsoft Corporation. To fulfill the scientific aim, three research questions are proposed and answered. [RQ1] Which Activision Blizzard firms were the leading drivers of patenting activity? [RQ2] What are the main technologies patented during the analyzed period? [RQ3] Which nodes, edges, and communities form the technological core of the IPC code co-occurrence network? The article comprises five sections, namely an introduction, a review of the literature, a description of the methodology and the data collection process, results of the network analysis, and conclusions of the study.

## 2. Literature review

### *2.1. History of Activision Blizzard*

Activision Blizzard itself was the result of the 2008 merger between Vivendi Games (who was the parent company of Blizzard Entertainment and a subsidiary of Vivendi) and Activision Publishing [16]. Vivendi divested its controlling share of Activision Blizzard in 2013 [17]. As such, Activision Blizzard became the global ultimate owner until 2023. The most significant business move during this period was the 2016 acquisition of King Digital Entertainment (King.com) for $5.9 billion ($18 per share) [18].

At the time of Microsoft's takeover of Activision Blizzard, the intellectual properties of the latter included World of Warcraft, Diablo, Hearthstone, StarCraft, Overwatch,

Battle.net, Candy Crush Saga, and Call of Duty [1], [7], [19], [20]. Originally released in 2004, World of Warcraft (WoW) is a subscription-based massively multiplayer online role-playing game (MMORPG). Together with its ten expansions, WoW is arguably the most successful and influential game of its genre. The subscribed count peaked at 12 million in 2010, near the end of the Wrath of the Lich King expansion [21]. Hearthstone is a free-to-play digital collectible card game with over 130 million players, set in the same universe as WoW [22]. Diablo 4, the latest in a series of genre-defining isometric action role-playing games, broke the record for the fastest-selling Blizzard game in June 2023 [23]. StarCraft games are real-time strategies that heavily influenced Blizzard's success in the early esports scene, especially in South Korea [24], [25]. Call of Duty series and its recent free-to-play mode Warzone are popular first-person shooters [26].

Battle.net is the iconic online multiplayer service of games published by Blizzard Entertainment. Originally launched in 1996, it has since evolved into a unified launcher and digital storefront that includes Blizzard Entertainment games and Activision Publishing games, like the Call of Duty franchise [20], [27]. Microsoft Gaming has its own PC launcher called the Xbox app, with similar capabilities to the Battle.net app. Historically, Blizzard Entertainment PC games have been released exclusively on Battle.net, however, its two high-profile games, namely Overwatch 2 and Diablo 4, have also been made available on the Steam store [28], [29].

Activision Blizzard was investigated for allegations of a hostile work environment and harassment. In 2021, the company was sued by the California Civil Rights Department. As of late 2023, the suit was still ongoing following the refusal to dismiss by the Second Appellate District of the Court of Appeal on October 22, 2022 [30]. A $54 million settlement was announced on December 15, 2023 [31]. A separate lawsuit by the U.S. Equal

Employment Opportunity Commission was settled for $18 million on March 30, 2022 [32]. An investigation by the U.S. Securities and Exchange Commission resulted in a $35 million penalty on February 2, 2023 [33]. Because of these issues, the firm has garnered substantial criticism from the media and the community [34], [35], [36], [37].

### *2.2. Academic interest in video games and Activision Blizzard*

Over the past few decades, the video game industry has become one of the largest in the world. As the average budgets of AAA video games grow larger and larger, it is no surprise that patents and intellectual property rights have become increasingly important. General patent-related scientific articles focusing on video games analyze microtransactions [38], patent internationalization [39], inter partes reviews [40], actor-network theory [41], [42], typology of patent and trademark strategies [43], and intellectual property law and influential cases [44].

Despite its importance to the video game industry, a standard title-abstract-keywords search in the Scopus database for “Activision Blizzard” and its three main subsidiaries returns only 31 scientific articles (TABLE 1). The number is significantly higher for its flagship franchises. For example, analogous searches for “World of Warcraft” and “StarCraft” return 380 and 185 articles, respectively. However, all these topics are dwarfed by “Microsoft” (23 thousand articles) and a generic search for “video game” (18 thousand articles).

TABLE 1 Academic interest in selected topics

| **Search query** | **Number of articles** |
|---|---|
| TITLE-ABS-KEY ( "activision blizzard" OR "blizzard entertainment" OR "activision publishing" OR "king digital entertainment" ) AND DOCTYPE ( ar OR re ) | 31 |
| TITLE-ABS-KEY ( "starcraft" ) AND DOCTYPE ( ar OR re ) | 185 |
| TITLE-ABS-KEY ( "playstation" ) AND DOCTYPE ( ar OR re ) | 278 |
| TITLE-ABS-KEY ( "world of warcraft" ) AND DOCTYPE ( ar OR re ) | 380 |
| TITLE-ABS-KEY ( "mmorpg" ) AND DOCTYPE ( ar OR re ) | 404 |

| Search query | Number of articles |
|---|---|
| TITLE-ABS-KEY ( "xbox" ) AND DOCTYPE ( ar OR re ) | 593 |
| TITLE-ABS-KEY ( "video game" ) AND DOCTYPE ( ar OR re ) | 18,146 |
| TITLE-ABS-KEY ( "microsoft" ) AND DOCTYPE ( ar OR re ) | 22,976 |

Source: Own elaboration based on Scopus [45].

Due to its longevity and large player base, WoW has been analyzed as the representative of the MMORPG genre. Influential early studies focus mostly on the nature of social interactions in its virtual world [46], [47]. Other research topics include online and real-life social relationships [48], escapism [49], gaming nostalgia [50], personality profiles of emerging adult players [51], community reaction to the short-lived Real ID forum requirements [52], ownership of in-game mods [53], virtual currency pricing [54], in-game goods and price convergence [55], drivers of purchase intention for virtual products [56], and models predicting subscription renewal [57]. In 2005, a virtual plague due to the unintended behavior of an in-game mechanic led to discussions regarding the applicability of the data to real-life epidemiological modeling [58]. Many of the recent highly cited scientific articles explore internet and gaming addiction [59], [60], [61]. On the other hand, the popularity of StarCraft in academic literature is driven primarily by the study of artificial intelligence, as the game is famous for its mechanical complexity [62], [63], [64], [65].

### *2.3. Patent analysis and IPC codes in academic literature*

The network analysis conducted in this article is based on a system of codes devised by the World Intellectual Property Organization (WIPO), which is one of the specialized agencies of the United Nations [66]. WIPO’s “International Patent Classification (IPC), established by the Strasbourg Agreement 1971, provides for a hierarchical system of language independent symbols for the classification of patents and utility models according to the different areas of technology to which they pertain” [14]. IPC codes are divided into sections (e.g., A), classes (e.g., A63), subclasses (e.g., A63F), main groups (e.g., A63F 13/00), and subgroups (e.g.,

A63F 13/55) [13]. Codes of lower levels are subdivisions of higher levels and thus can be aggregated if necessary (e.g., subgroups A63F 13/55 and A63F 13/70 can be aggregated to their subclass A63F). At the subgroup level, there is a further subdivision according to the number of dots (e.g., the one-dot subgroup A63F 13/70 is hierarchically superior to the two-dot subgroup A63F 13/71) [13]. Additionally, subsections are headings that are placed between sections and classes; subsections do not have their own IPC codes and are merely informative. The names and descriptions of individual IPC codes are available on WIPO's website [14].

While the methodology of IPC code co-occurrence networks remains relatively new, it has a strong theoretical background because of the similarities to many established bibliometric methods. The closest one is the analysis of keywords assigned to scientific articles by their authors. Such networks, known as keyword co-occurrence networks, are a popular method of knowledge mapping and research hotspot identification that can easily be applied to large groups of scientific articles [15]. Keyword co-occurrence networks have been used in the bibliometric analyses of the video game industry [67], [68], [69], [70], [71]. Similarly, IPC code co-occurrence can be applied to patent data.

Networks based on patent classifications are commonly used in the mapping of technology. IPC code co-occurrence networks (also known as co-classification maps) can be traced to the seminal article of Engelsman and van Raan [72]. Their pioneering co-classification map is based on 28 technology fields (delineated by specific IPC codes). It is the first co-classification study based on IPC codes; the authors reference analyses of the co-classification of science fields based on scientific articles [73], [74] and other bibliometric methods. The primary uses of co-classification maps are the static and dynamic analyses of: technology as a whole, bridging or interdisciplinary technologies, technological peripheries, and technological centers of innovative activities [72, p. 25]. Transfer of knowledge occurs through bridging codes connecting interdisciplinary technologies.

Another significant development in the literature is Breschi et al. [75], who study the knowledge-relatedness of firms using 30 technology fields (and IPC codes assigned to these fields). Their model disregards the order in which IPC codes are assigned to a patent. The notion of knowledge-relatedness comprises knowledge proximity (intended and unintended learning), knowledge commonalities (economies of scope by utilizing knowledge in more than one technology), and knowledge complementarities (the necessity to utilize different technologies) [75, pp. 70–71]. The findings show that firms' technological diversification and new patents cluster coherently (non-randomly) around related technologies. In other words, a firm learns or develops new technologies in closely connected fields. Firms can use technologies that are present in their knowledge base as departure points to new technological clusters.

Patent maps simplify the presentation of complex datasets and provide a holistic and pictorial overview of the most relevant aspects of analyzed technologies by filtering out the noise [72]. The applicability of network visualizations extends beyond academic analysis and has implications for public policy and planning. "In general, such a map of technology fields will be useful to aid in technology road mapping of innovation agents or technology-based industries and forecasting development directions of emerging technologies" [76, p. 2].

As shown by previous studies, IPC co-occurrence networks can be constructed based on the IPC classes [76], [77], [78], subclasses [77], [79], [80], [81], [82], [83], main groups [83], [84], [85], and subgroups [83]. Alternatively, code aggregation is also possible using IPC subsections [86] or custom categories defined by the authors (e.g., technology fields) [72], [75]. Although the process is less straightforward, a network can be constructed using a combination of codes at multiple levels; e.g., codes at class, subclass, main group, and subgroup levels analyzed in a single network [87]. The choice of the hierarchical level of IPC codes depends on the requirements of a specific study and can be influenced by factors such as the necessary technological details, the readability of network visualizations, and the number of analyzed patents.

## 3. Methodology and data

### *3.1. Co-occurrence networks*

When multiple IPC codes (or keywords) are assigned to a patent (or a scientific article) they are said to be co-occurring. Essentially, the language agnostic IPC codes can be analyzed in the same way as a standardized list of keywords. These intricate relationships can be modeled as undirected networks (graphs) consisting of nodes (vertices) and edges (links). The hierarchical IPC codes can be analyzed at different levels (e.g., class or subgroup).

In this article, four IPC code co-occurrence networks (co-classification maps) are constructed based on Activision Blizzard patents: two modularity networks (subclass and group) and two minimum spanning tree (MST) networks (subclass and group). Two modularity networks are the primary method of analysis, and two MST networks provide an alternative layout for network visualization. Nodes represent IPC codes (either at the subclass or group level), and edges that connect them represent the co-occurrence of IPC codes in a patent (Jaccard index value or its inverse). The choice of the subclass and group level (i.e., main groups and subgroups) is motivated by a relatively small number of patents in the analyzed dataset and their narrow focus. IPC codes are included in the network if they occur in at least two patents. Edges are weighted according to the number of patents where a pair of IPC codes co-occur. The order of codes in a patent is irrelevant since the edges are undirected. Furthermore, edge weights are normalized using the Jaccard index [76], [84]:

$$R_{ij} = \frac{|N_i \cap N_j|}{|N_i \cup N_j|} = \frac{N_{ij}}{N_i + N_j - N_{ij}}$$

where $R_{ij}$ is the Jaccard index between codes $i$ and $j$, $N_i$ is the number of patents of code $i$, $N_j$ is the number of patents of code $j$, $N_{ij}$ is the number of patents where codes $i$ and $j$ co-occur, and $0 < R_{ij} \leq 1$ (the edge does not exist if $R_{ij} = 0$). For modularity networks, edges are filtered using a threshold value ($R_{ij} \geq 0.05$). The inverse Jaccard index is used for MST networks ($1/R_{ij}$).

### *3.2. Dataset and software*

This article has an online dataset, which includes detailed results, a list of permalinks to analyzed patents, lists of nodes (IPC codes) and edges (co-occurrence), and high-quality color graphs [88].

The patent data (patent applications) were downloaded from the WIPO Patentscope database [89]. The advanced search query used for this purpose is *( PA:( "activision blizzard" OR "activision publishing" OR "blizzard entertainment" OR "king.com" OR "activision shanghai" OR "beenox" OR "blizzard albany" OR "demonware" OR "digital legends entertainment" OR "high moon studios" OR "infinity ward" OR "neversoft entertainment" OR "raven software" OR "redoctane" OR "sledgehammer games" OR "solid state studios" OR "toys for bob" OR "treyarch" OR "vicarious visions" ) AND DP:[2008 TO 2023] )*. The query includes patents filed by Activision Blizzard, its three main subsidiaries (including King.com patents that preceded the acquisition by Activision Blizzard), and other smaller subsidiaries (including now defunct studios) between 2008 and 2023. Using the publication date instead of the application date increases the reproducibility of the dataset. The corporate structure of Microsoft and Activision Blizzard is based on the Orbis database [90] and Activision's official website [91].

For the construction of the networks, IPC codes are imported as keywords to the bibliometric software VOSviewer [92] and analyzed using the network analysis software Gephi [93]. Gephi's ForceAtlas2 is used for graph layout [94], MSTs are based on Kruskal's algorithm [95], betweenness centrality is based on Brandes [96], and modularity class is based on Blondel et al. [97] and Lambiotte et al. [98].

### *3.3. Limitations of the study*

Patents focus on technical innovation and cannot reflect the artistic value of many video games. Co-classification maps are only as reliable as the patent classifications they are based on. Some innovations are not patented and patenting activities differ between firms, industries, and countries [72]. Furthermore, IPC co-occurrence is based strictly on IPC codes and not on abstracts or full texts of the patents. Any potential shortcomings of the data, like imprecisely assigned IPC codes, would be reflected in the patent maps. The same is true for most bibliometric methods that rely on standardized data. An alternative or a supplement to co-classification maps are co-word networks based on text mining of patent data. These networks, in turn, rely on the quality and assumptions of the chosen natural language processing model. Lastly, patent data cannot be used to analyze the innovative activities of firms without any patent applications, and many small firms (e.g., indie video game developers) unfortunately fall into this category.

## 4. Results

### *4.1. Overview of patents and corporate structure*

Microsoft's activities are divided into three segments: Productivity and Business Processes (e.g., Office 365 and LinkedIn), Intelligent Cloud (e.g., Azure and SQL Server), and More Personal Computing (e.g., Windows and Xbox) [8], [99]. TABLE 2 lists the 15 largest subsidiaries of Microsoft Corporation (Activision Blizzard's global ultimate owner) by operating revenue according to the Orbis database. Activision Blizzard is the sixth largest subsidiary in the Microsoft corporate group. With operating revenue of $7.5 billion, Activision Blizzard is the largest subsidiary located in the United States, ahead of LinkedIn. Five of the largest subsidiaries are located in Ireland and two of them are classified as financial. Microsoft Corporation's operating revenue was $212 billion and its total assets were $412 billion (these values include all its subsidiaries).

TABLE 2 Microsoft Corporation and its largest subsidiaries (most recent year available)

| Name | Country | City | Type | Operating revenue (million USD) | Total assets (million USD) |
|---|---|---|---|---|---|
| MICROSOFT CORPORATION | US | REDMOND | Corporate | 211,915 | 411,976 |
| MICROSOFT IRELAND OPERATIONS LIMITED | IE | DUBLIN 2 | Corporate | 64,913 | 45,103 |
| MICROSOFT IRELAND INVESTMENTS UNLIMITED COMPANY | IE | DUBLIN 2 | Financial | 54,392 | 12,409 |
| MICROSOFT IRELAND RESEARCH UNLIMITED COMPANY | IE | DUBLIN 2 | Corporate | 47,787 | 76,664 |
| MICROSOFT SINGAPORE HOLDINGS PTE. LTD. | SG | SINGAPORE | Corporate | 22,350 | 22,350 |
| MICROSOFT LIMITED | GB | READING | Corporate | 7,609 | 5,662 |
| ACTIVISION BLIZZARD, INC. | US | SANTA MONICA | Corporate | 7,528 | 27,383 |
| MICROSOFT JAPAN CO., LTD. | JP | MINATO-KU | Corporate | 6,483 | .. |
| LINKEDIN IRELAND UNLIMITED COMPANY | IE | DUBLIN 2 | Corporate | 5,287 | 3,177 |
| MICROSOFT PTY LTD | AU | SYDNEY | Corporate | 4,985 | 2,999 |
| MICROSOFT REGIONAL SALES PTE. LTD. | SG | SINGAPORE | Corporate | 4,574 | 2,534 |
| MICROSOFT FRANCE | FR | ISSY LES MOULINEAUX | Corporate | 3,317 | 2,833 |
| LINKEDIN CORPORATION | US | MOUNTAIN VIEW | Corporate | 2,990 | 7,011 |
| MICROSOFT (CHINA) CO., LTD. | CN | BEIJING | Corporate | 2,688 | 1,855 |
| MICROSOFT CORPORATION (INDIA) PVT LTD | IN | GURUGRAM | Corporate | 2,342 | 3,123 |
| MICROSOFT GLOBAL FINANCE UNLIMITED COMPANY | IE | DUBLIN 2 | Financial | 2,241 | 84,991 |

Source: Own elaboration based on Orbis [90].

The WIPO Patentscope search query identified 612 Activision Blizzard patent applications. The patents in the dataset were filed by Activision Publishing (317), King.com

(269), and Blizzard Entertainment (26). Other subsidiaries did not file any patents in the analyzed period, however, both Vicarious Visions and Treyarch applied for patents before 2008. An overwhelming majority of Activision Blizzard patents were filed in the United States (544); of the rest, 63 were international patents filed under the Patent Cooperation Treaty, three were filed in the European Patent Office, one was filed in Japan, and one was filed in Brazil. The innovative activities of Activision Blizzard, as evidenced by its patenting pattern, were focused on the firm's domestic market.

GRAPH 1 illustrates the annual number of new Activision Blizzard patents by publication date. In 2008, only three patents were published, by far the lowest number in the time series. Between 2009 and 2013, the annual number of patents ranged between 10 and 21, and then it jumped to 73 patents in 2014. In the decade since its 2014 peak, annual patents remained relatively steady, with a slight decline in 2023 (42 patents).

GRAPH 1 Annual number of new Activision Blizzard patents (publication date)

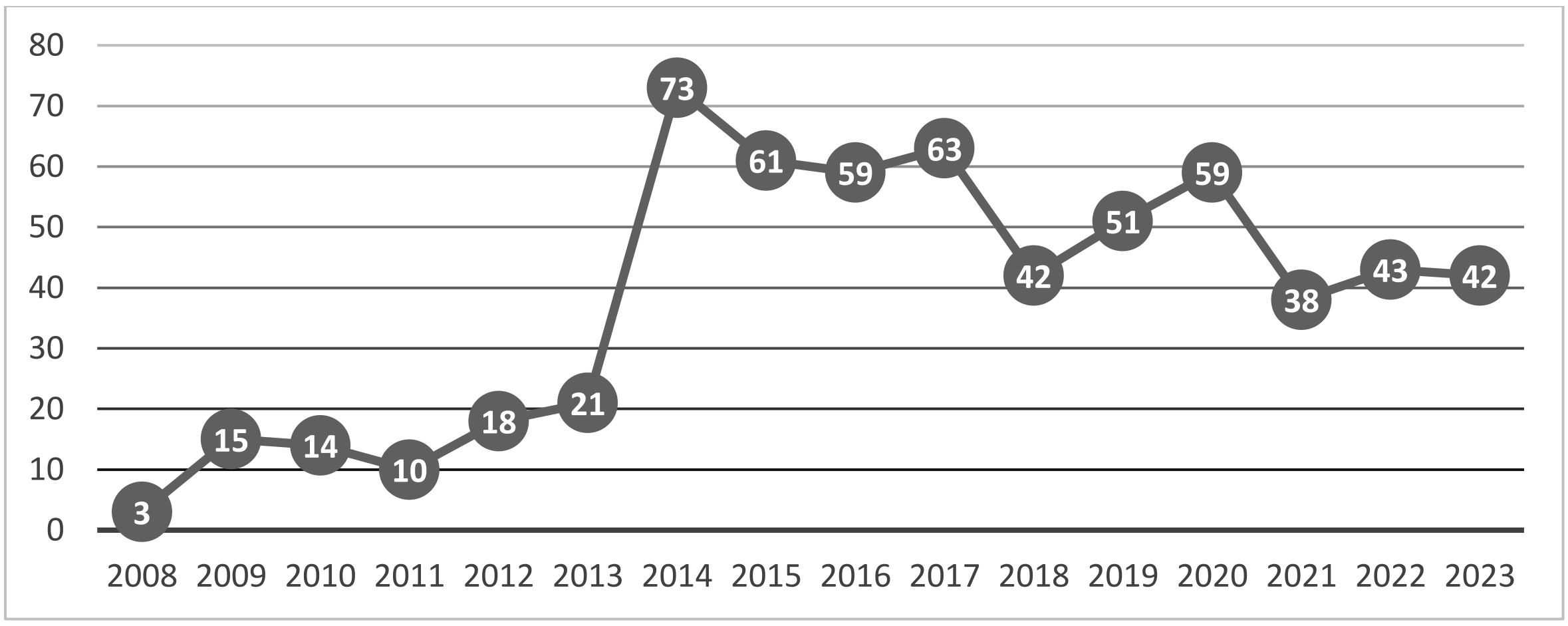

Source: Own elaboration based on Patentscope [89].

### *4.2. Subclass networks*

TABLE 3 presents the network statistics of the ten IPC code subclasses assigned to the largest number of patents. The most prolific IPC code is the general subclass of card, board, and video games (A63F), which is assigned to 485 patents (79% of total patents). The second most popular subclass, with 177 patents, is electric digital data processing (G06F). Other popular subclasses, in order of the number of patents, include coin-freed apparatus (G07F), image data processing or generation (G06T), information and communication technology (ICT) specially adapted for commercial/financial purposes (G06Q), transmission of digital information (H04L), apparatus for physical training (A63B), computing arrangements (G06N), graphical data reading and presentation (G06K), and pictorial communication (H04N).

TABLE 3 Network statistics of prolific IPC codes (subclass level)

| IPC code | Occurrences | Modularity network | | | MST network | |
|---|---|---|---|---|---|---|
| | | Modularity class | Degree | Betweenness | Degree | Betweenness |
| **A63F**: card, board, and video games | 485 | 1 | 3 | 0.073 | 1 | 0 |
| **G06F**: electric digital data processing | 177 | 1 | 5 | 0.200 | 5 | 0.673 |
| **G07F**: coin-freed apparatus | 65 | 1 | 3 | 0.012 | 3 | 0.368 |
| **G06T**: image data processing or generation | 52 | 2 | 4 | 0.076 | 1 | 0 |
| **G06Q**: ICT specially adapted for commercial or financial purposes | 35 | 1 | 3 | 0.123 | 2 | 0.105 |
| **H04L**: transmission of digital information | 33 | 5 | 3 | 0.188 | 3 | 0.608 |
| **A63B**: apparatus for physical training | 20 | 1 | 2 | 0 | 1 | 0 |
| **G06N**: computing arrangements | 18 | 2 | 3 | 0.102 | 2 | 0.456 |
| **G06K**: graphical data reading and presentation | 17 | 2 | 6 | 0.129 | 3 | 0.205 |
| **H04N**: pictorial communication | 16 | 2 | 3 | 0.013 | 3 | 0.444 |

Source: Own elaboration based on Patentscope [89].

GRAPH 2 shows the co-occurrence between technologies represented by IPC codes (subclass level). After filtering procedure ($R_{ij} \geq 0.05$), the subclass modularity network has 20 nodes and 26 edges. The network has five modularity classes (communities) and three components. The size of the nodes is determined by the number of patents (occurrence count). G06K has the highest degree (6), followed by G06F (5), and G06T and G10L (4). In addition, G06K also has the highest betweenness (0.200). The network centrality scores of the node representing the most prolific IPC subclass, A63F, are relatively weak. On the other hand, despite its high degree centrality, G10L is absent from TABLE 3, as it is assigned to only four patents; this subclass represents speech analysis or synthesis. Prolific network edges between nodes represent bridges that connect technologies. The most important edges in the modularity network are the links between A63B and G07F (19 patents, 0.288 Jaccard index), between A63F and G06F (115 patents, 0.210 Jaccard index), and between G06K and G06T (11 patents, 0.190 Jaccard index). These edges form the core connections of the two largest communities: orange (modularity class 1; prolific nodes: A63F, G06F, and G07F) and blue (modularity class 2; prolific nodes: G06T, G06N, and G06K).

GRAPH 2 IPC code co-occurrence modularity network (subclass level, color=modularity class, size=occurrence count)

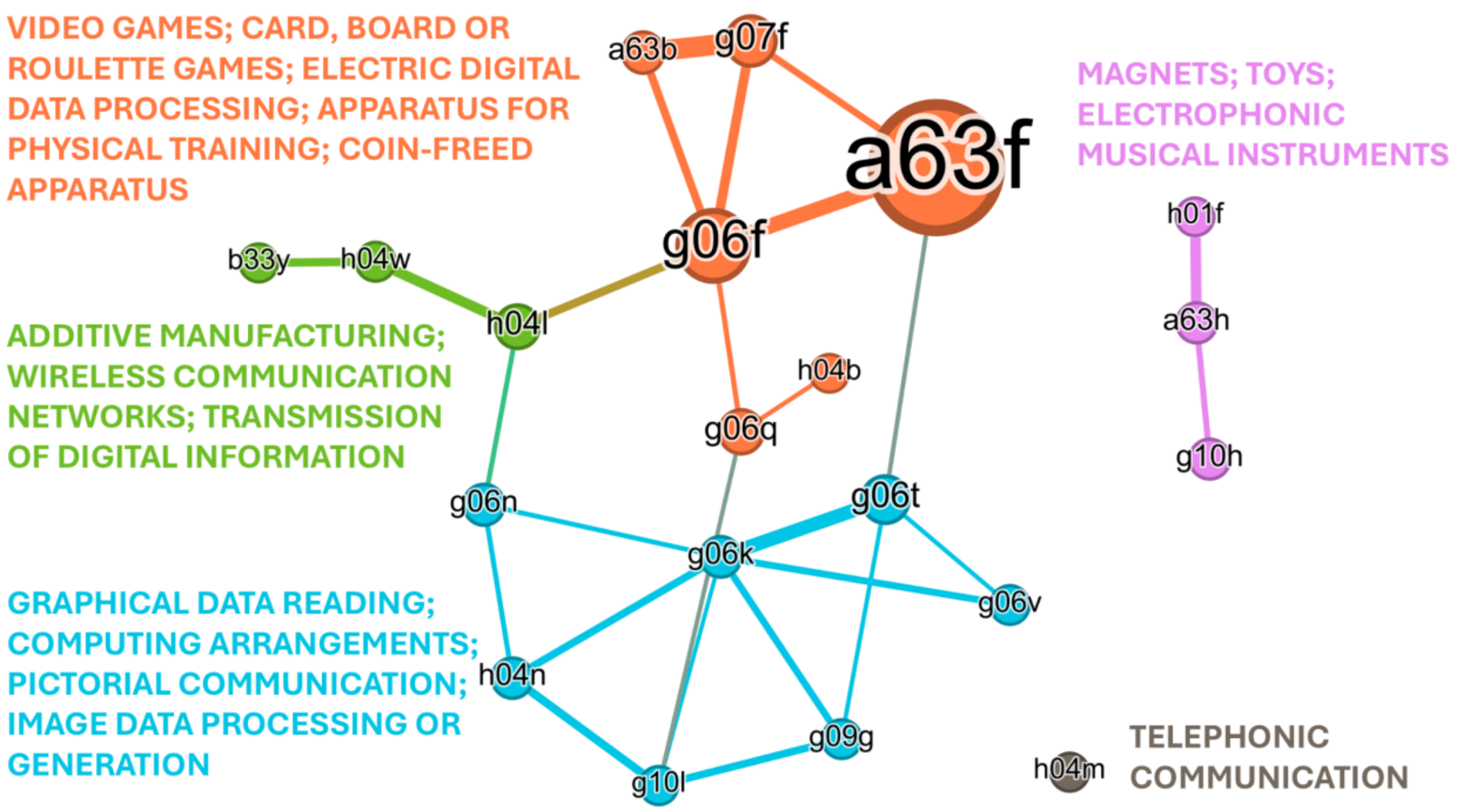

Source: Own elaboration based on Patentscope [89].

The MST network (GRAPH 3) has a single component with 20 nodes and 19 edges. The four nodes marked in red represent prolific IPC codes (subclasses) that occur in at least 50 different patents. G06F is the most central node in the MST network with the highest degree centrality (5) and betweenness centrality (0.673). Three of the four prolific codes are grouped together, and one (G06T) is a leaf (node with a single connection) on the other side of the tree.

GRAPH 3 IPC code co-occurrence MST network (subclass level, red=prolific IPC code)

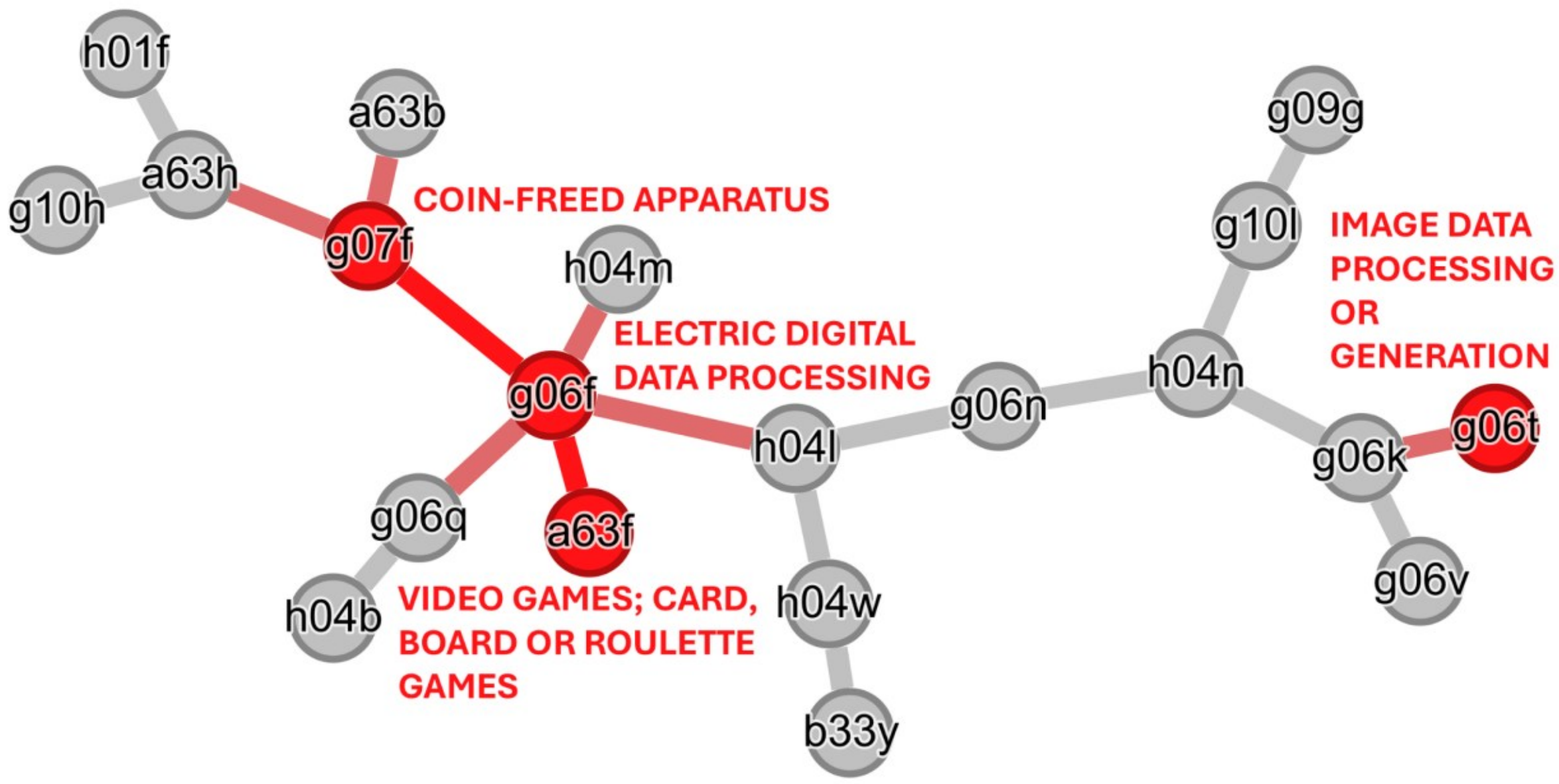

Source: Own elaboration based on Patentscope [89].

### *4.3. Group networks*

TABLE 4 presents the network statistics of the twenty IPC code groups (i.e., main groups and subgroups) assigned to the largest number of patents. The most prolific IPC code is the main group of video games (A63F 13/00), which is assigned to 121 patents (20% of total patents). The second most prolific IPC code is games using electronic circuits not otherwise provided for (A63F 9/24); with 79 patents this is the most popular subgroup. Next are the two subgroups each assigned to 78 patents: game servers-devices interconnection arrangements (A63F 13/30) and generating/modifying game content by enabling/updating specific game elements (A63F 13/69). The fifth most prolific IPC code, assigned to 72 patents, describes controlling the output signals based on the game progress involving additional visual information provided using indicators, e.g., a heads-up display (A63F 13/537). Next on the list are the subgroups comprising coin-freed apparatus for games, toys, sports, or amusements (G07F 17/32) and details of game servers (A63F 13/35), with 65 patents for the former and 64 patents for the latter. The final group with 50+ patents is the subgroup that describes controlling the output signals based on the game progress involving aspects of the displayed game scene (A63F 13/52), which is assigned to 53 patents.

Other than the above-mentioned 50+ patents IPC codes, there are 12 groups (one main group and 11 subgroups) in TABLE 4 whose occurrence counts range between 28 and 48 patents. These include in order of the number of patents, codes associated with processing input control signals of video game devices (A63F 13/40), game security/management aspects involving player-related data (A63F 13/79), game security/management aspects involving player-related data for finding other players (A63F 13/795), output arrangements for video game devices (A63F 13/25), special adaptations for executing a specific game genre/mode (A63F 13/80), input arrangements for video game devices like touch screens (A63F 13/2145), digital computing or data processing equipment/methods (G06F 17/00), authoring tools specially adapted for game development or game-integrated level editor (A63F 13/60), game servers-devices interconnection arrangements using Internet (A63F 13/335), computing the game score (A63F 13/46), generating/modifying game content adaptively or by learning from player actions (A63F 13/67), and arrangements for executing specific programs (G06F 9/44).

TABLE 4 Network statistics of prolific IPC codes (group level)

| IPC code | Occurrences | Modularity network | | | MST network | |
|---|---|---|---|---|---|---|
| | | Modularity class | Degree | Betweenness | Degree | Betweenness |
| **A63F 13/00**: main group of video games | 121 | 3 | 33 | 0.009 | 3 | 0.060 |

| IPC code | Occurrences | Modularity network | | | MST network | |
|---|---|---|---|---|---|---|
| | | Modularity class | Degree | Betweenness | Degree | Betweenness |
| **A63F 9/24**: games using electronic circuits not otherwise provided for | 79 | 3 | 25 | 0.007 | 2 | 0.043 |
| **A63F 13/30**: game servers-devices interconnection arrangements | 78 | 3 | 28 | 0.006 | 1 | 0 |
| **A63F 13/69**: generating or modifying game content by enabling or updating specific game elements | 78 | 3 | 27 | 0.015 | 1 | 0 |
| **A63F 13/537**: controlling the output signals based on the game progress involving additional visual information provided using indicators, e.g., a heads-up display | 72 | 3 | 25 | 0.013 | 2 | 0.009 |
| **G07F 17/32**: coin-freed apparatus for games, toys, sports, or amusements | 65 | 3 | 29 | 0.004 | 2 | 0.009 |
| **A63F 13/35**: details of game servers | 64 | 3 | 29 | 0.012 | 4 | 0.052 |
| **A63F 13/52**: controlling the output signals based on the game progress involving aspects of the displayed game scene | 53 | 3 | 31 | 0.027 | 1 | 0 |
| **A63F 13/40**: processing input control signals of video game devices | 48 | 3 | 29 | 0.009 | 2 | 0.068 |
| **A63F 13/79**: game security or management aspects involving player-related data | 45 | 3 | 31 | 0.017 | 1 | 0 |
| **A63F 13/795**: game security or management aspects involving player-related data for finding other players | 43 | 3 | 24 | 0.007 | 3 | 0.018 |
| **A63F 13/25**: output arrangements for video game devices | 39 | 3 | 33 | 0.014 | 2 | 0.327 |
| **A63F 13/80**: special adaptations for executing a specific game genre or mode | 38 | 3 | 28 | 0.006 | 1 | 0 |
| **A63F 13/2145**: input arrangements for video game devices like touch screens | 37 | 3 | 36 | 0.026 | 4 | 0.282 |
| **G06F 17/00**: digital computing or data processing equipment or methods | 35 | 3 | 25 | 0.016 | 2 | 0.035 |
| **A63F 13/60**: authoring tools specially adapted for game development or game-integrated level editor | 32 | 3 | 22 | 0.024 | 1 | 0 |
| **A63F 13/335**: game servers-devices interconnection arrangements using Internet | 31 | 3 | 32 | 0.043 | 3 | 0.333 |
| **A63F 13/46**: computing the game score | 30 | 3 | 27 | 0.007 | 2 | 0.043 |
| **A63F 13/67**: generating or modifying game content adaptively or by learning from player actions | 29 | 3 | 15 | 0.005 | 1 | 0 |
| **G06F 9/44**: arrangements for executing specific programs | 28 | 3 | 29 | 0.018 | 3 | 0.093 |

Source: Own elaboration based on Patentscope [89].

GRAPH 4 shows the co-occurrence between technologies represented by IPC codes (group level). After filtering procedure ($R_{ij} \geq 0.05$), the group modularity network has 224 nodes and 1,312 edges. The network has 16 modularity classes (communities) and four components. The size of the nodes is determined by the number of patents (occurrence count). Labels are enabled for nodes with at least 50 different patents. The most prolific codes (50+ patents) all belong to the same modularity class. This modularity class (colored pink in GRAPH 4) forms the technological core of the analyzed firm. In terms of co-occurrence counts between prolific IPC codes listed in TABLE 4, we can observe the links between A63F 13/00 and A63F 9/24 (36 patents, 0.220 Jaccard index), between A63F 13/30 and G07F 17/32 (35 patents, 0.324 Jaccard index), and between A63F 13/00 and A63F 13/30 (35 patents, 0.213 Jaccard index). Due to Jaccard index normalization, the 25 most influential edges (1.000 Jaccard index) in the modularity network connect nodes with relatively small occurrence counts (four patents or fewer). All edges with perfect Jaccard index are listed in the online dataset, examples include links between concurrent/sequential virtual cameras and server/clients network processes for video distribution (A63F 13/5252 and H04N 21/63), between game devices/servers secure communication and protecting data integrity (A63F 13/71 and G06F 21/64), and between toy imitations of apparatus and selecting circuits of electrophonic musical instruments (A63H 33/30 and G10H 1/18).

GRAPH 4 IPC code co-occurrence modularity network (group level, color=modularity class, size=occurrence count)

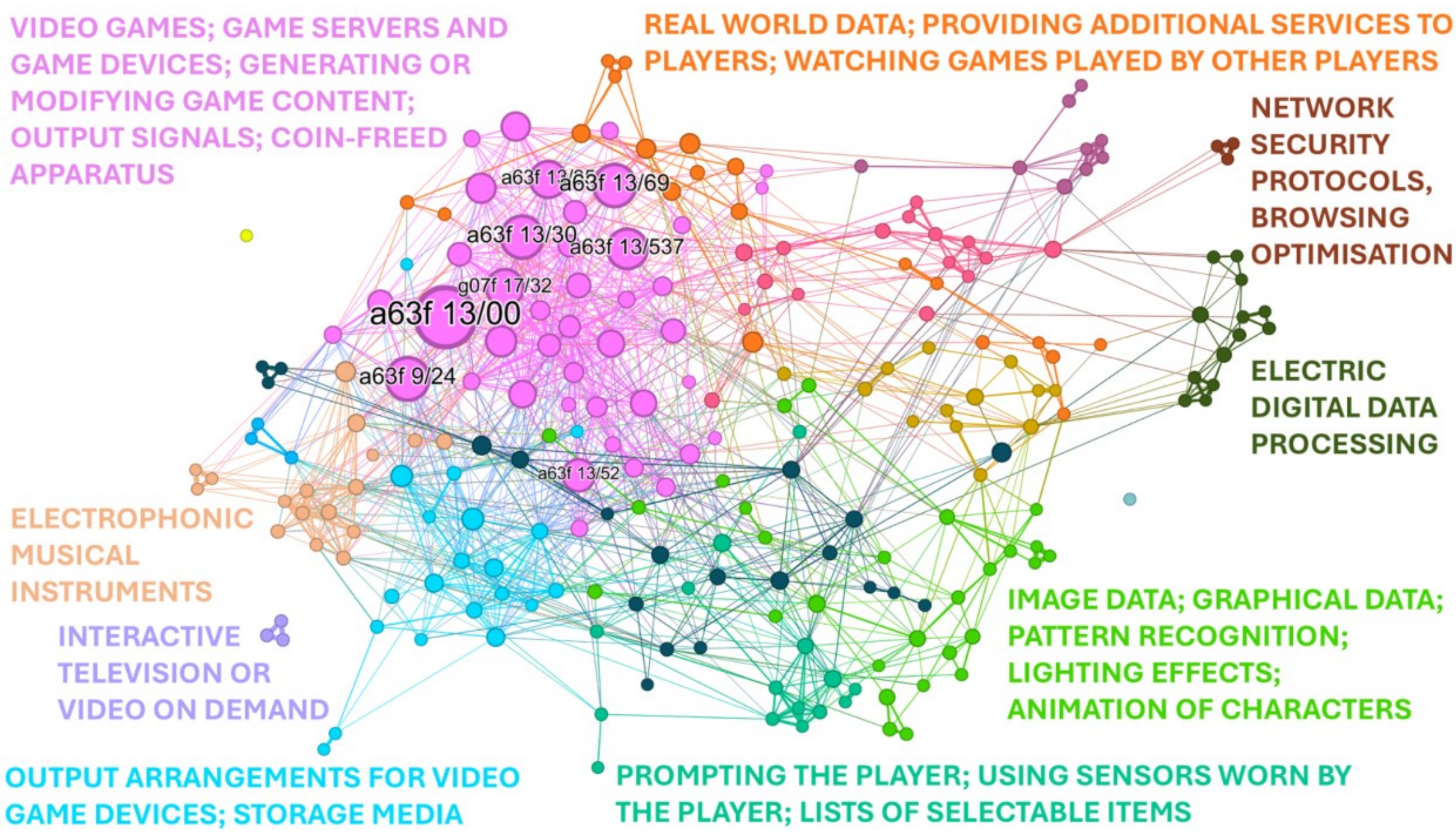

Source: Own elaboration based on Patentscope [89].

The MST network (GRAPH 5) consists of three components; the main component has 220 nodes and 219 edges, the second largest component has three nodes and two edges, and the smallest component has a single unconnected node. Due to multiple MST components, this network is technically a minimum spanning forest. The eight nodes marked in red represent prolific IPC codes (groups) that occur in at least 50 different patents. Most of the prolific nodes are relatively close together; one of them (A63F 13/52) is a leaf on the other side of the tree. There are four nodes in this network tied for the highest degree (5): pattern recognition (G06K 9/00), interaction techniques based on graphical user interfaces for the control of specific functions (G06F 3/0484), output arrangements for video game devices for affecting ambient conditions (A63F 13/28), and depth/shape recovery for image analysis (G06T 7/50). The node with the highest betweenness centrality is A63F 13/428 (0.608); this code occurs in 18 patents and represents game commands involving motion or position input signals, e.g., rotation of a controller/arm sensed by accelerometers/gyroscopes.

GRAPH 5 IPC code co-occurrence MST network (group level, red=prolific IPC code)

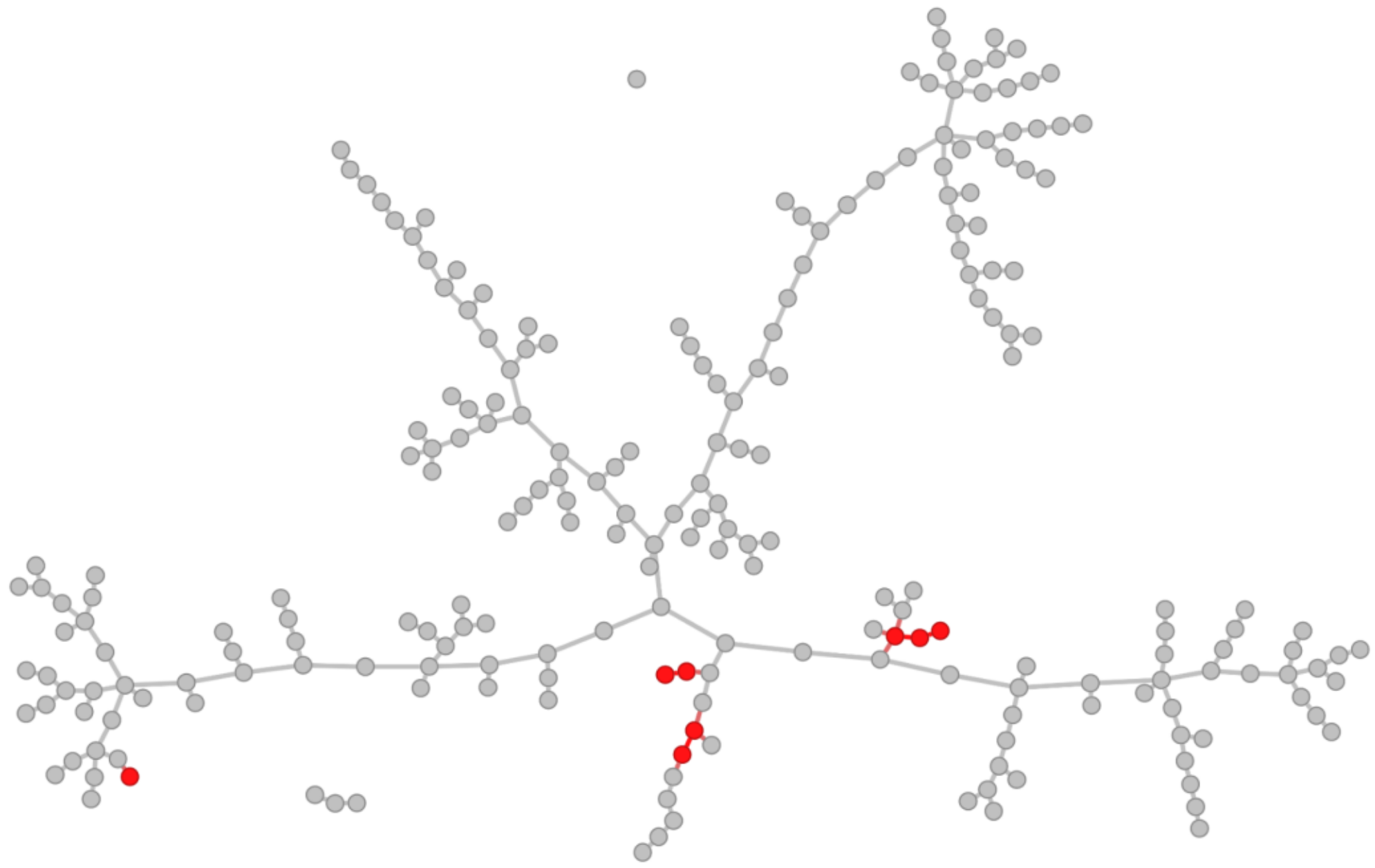

Source: Own elaboration based on Patentscope [89].

## 5. Conclusions and discussion

### *5.1. Implications for the video game industry*

Microsoft's acquisition of Activision Blizzard was undoubtedly one of the most significant developments in the history of the video game industry. Its importance reverberates beyond the video game industry, as it remains among the top tech acquisitions of all time. Activision Blizzard immediately became one of the most important subsidiaries in the Microsoft corporate group, and the largest one located in the United States. Between the Xbox and Microsoft Gaming brands, and other high-profile acquisitions (e.g., Bethesda Softworks and Obsidian Entertainment), Microsoft has consolidated much of the American gaming industry. Consequently, Microsoft now controls both the Xbox app and Battle.net app, and there is a significant overlap between the two. Battle.net store has already made several games published by Xbox Game Studios available for purchase. Examples include Avowed and The Outer Worlds 2 by Obsidian Entertainment, and Sea of Thieves by Rare. On the other hand, Diablo 4 has been listed on the Xbox Game Pass.

Patents are important for the video game industry. Due to the increasing globalization of the industry, the key firms have become multinational enterprises (MNEs). MNEs are firms that have engaged in foreign direct investment (FDI) in at least one country. Because the development of AAA titles is so expensive, the increasingly oligopolistic role of the largest MNEs (e.g., Microsoft, Tencent, Sony, and Nintendo) pushes for further market consolidation. Patents held by a firm can be crucial for its valuation during the acquisition process. The internationalization of the production of modern video games adds another layer of complexity. To protect their intellectual property, MNEs can patent the technologies and internalize market transactions using wholly owned subsidiaries.

IPC co-occurrence networks provide an easily digestible overview of technological profiles of patenting firms. There are limitations of the patent approach, listed in the previous section of the article, but the key is the inability to patent artistic expression. Because of their technical nature, patents will be of interest mostly to large video game developers and publishers, producers of hardware, and firms that create innovative technological software solutions. Many video game firms, especially indie developers, do not patent at all. Nevertheless, they should still be aware of the patenting activity of their competitors.

While patents are not something that often reaches the gaming news, there have been situations where the patented gameplay mechanics became relevant to wider audiences. For example, Warner Bros. Entertainment successfully patented the Nemesis system from games Middle-earth: Shadow of Mordor and Middle-earth: Shadow of War, both by now defunct Monolith Productions [100]. It essentially blocked the development of similar systems in other video games. Another example is the highly publicized lawsuit between Nintendo and Pocketpair, developers of Palworld [101], [102]. The lawsuit is based on three patents granted to Nintendo and The Pokémon Company by the Japan Patent Office in 2024 [103], [104], [105]. This exemplifies the fact that even if video game developers are not interested in patenting, they need to be wary of infringing on patents of other firms.

### *5.2. Overview of patenting activity*

The 612 Activision Blizzard patents in the analyzed dataset (2008-2023) were filed by Activision Publishing (317), King.com (269), and Blizzard Entertainment (26). Unsurprisingly, the most prolific IPC code is the general subclass of card, board, and video games (A63F) assigned to 79% of total patents. Consequently, 21% of total patents are not explicitly related to video games. Other popular subclasses, in order of the number of patents, include electric digital data processing (G06F), coin-freed apparatus (G07F), and image data processing or generation (G06T). In terms of the more detailed group-level network, the innovative activities focus primarily on the main group of video games (A63F 13/00) and other groups related to electronic games, output signals, input signals, game content, authoring tools, adaptive content generation, coin-freed (payment/token activated) apparatus, game servers, matchmaking, and data security.

Influential edges serve as bridges between technologies. The three notable pairs of connected technologies at the subclass level are apparatus for physical training and coin-freed apparatus (A63B and G07F), video games and electric digital data processing (A63F and G06F), and graphical data reading/presentation and image data processing/generation (G06K and G06T). At the group level, the network has 25 edges with a perfect Jaccard index and relatively small co-occurrences (listed in the online dataset).

Looking at the list of intellectual properties of Activision Blizzard, its major franchises are dependent on technologies tied to online multiplayer gaming and have stringent requirements regarding server stability, security, and efficient matchmaking. Since the early days of Battle.net and Diablo, Blizzard's video games have pushed the technological frontier of online gaming. Popular titles with esports communities include WoW, Warcraft 3, Hearthstone, StarCraft 2, Overwatch 2, and Call of Duty: Black Ops 6. As an epitome of the MMORPG genre and Blizzard's most important game, WoW has required constant capital investment and technological support over its long lifespan. Activision Blizzard's most ambitious esports project was probably the prohibitively expensive and now-defunct Overwatch League. As shown in the analyzed networks, subgroups linked to game servers (e.g., A63F 13/30 and A63F 13/35) are ubiquitous in the dataset.

While the analysis of individual patents extends beyond the scope of this analysis, some warrant mentioning. Cross-realm zones, a staple of modern WoW, have been patented by Blizzard [106]. These zones are designed to avoid overcrowding at certain popular areas and foster player interaction in areas with low population, simultaneously helping both high-population and low-population WoW realms. A patent relevant to multiplayer matchmaking is the techniques of ranking players' skill level based on internal and external scores [107]. Such techniques help to avoid lopsided matches and are an important component of many competitive multiplayer games, where the internal matchmaking values are hidden from players and accurate, while the external matchmaking values are displayed to players and seasonally reset. Another patent describes the technical process for offering synergistic bundles of play-items (e.g., Hearthstone cards) based on pairwise synergy scores [108].

Other interesting patents include Activision's video game controllers and technologies relevant to the Guitar Hero franchise [109], [110], [111] and King.com's many touch screen technologies relevant to mobile games like Candy Crush Saga [112], [113], [114]. Due to esports, there are multiple patents related to video streaming, replays, and gameplay information. These comprise the collective voting mechanism of spectators to determine gameplay events [115], [116], selective incorporation of audience feedback [117], and automatic annotation of gameplay for esports tournaments [118]. Subgroup coin-freed apparatus for games (G07F 17/32) is associated with the technological requirements of monetization of video games, including gift cards for virtual goods [119], matchmaking based on microtransactions [120], and offline purchases [121].

The analyzed patents show that the acquisition is likely to generate strong synergies with Microsoft's existing intellectual properties. It also brings new challenges and opportunities, like controlling two popular and competing digital storefronts. It remains to be seen whether the ultimate long-term impact on the industry will be beneficial or not.

### *5.3. Implications for patent network analysis*

The results show that IPC code co-occurrence can be successfully used in the analysis of the video game industry. In terms of the general applicability of the methodology, such networks can be used to map the technological potential of countries, industries, and firms. Technological profiles can be constructed for individual firms or comparisons between firms or groups of firms. They can be utilized in macroeconomics and case study analysis to quantify the current relatedness between technologies, potential future developments, and technological trajectories. The results can be used to identify a list of patents for further qualitative study. Co-classification maps based on IPC codes can also be analyzed as examples of complex networks or hypergraphs. Finally, they provide a pleasing visual overview of the body of technological knowledge. Future extensions of this article might incorporate patent data of the entire industry, compare technological profiles of multiple firms, include the change over time aspect of dynamic networks, and model the connections using hypergraphs in lieu of traditional graphs.